\documentclass[manuscript]{acmart}
\usepackage{multirow}
\AtBeginDocument{%
  \providecommand\BibTeX{{%
    \normalfont B\kern-0.5em{\scshape i\kern-0.25em b}\kern-0.8em\TeX}}}

\setcopyright{acmcopyright}

\acmConference[Senior Thesis]{2023}

\begin{document}

\begin{titlepage}
    \begin{center}
        \vspace*{1cm}
            
        \Huge
        \textbf{Senior Thesis}
            
        \vspace{0.5cm}
        \LARGE
        Attitudes and perceived effectiveness among first-time online instructors during Covid-19
            
        \vspace{1.5cm}
            
        \textbf{Owen Xingjian Zhang}

        \vspace{1cm}
        
        Collaborator: Tiffany Wenting Li
        \vfill
        
        Supervisor: Prof. Karrie Karahalios
            
        \vspace{0.8cm}
            
        \includegraphics[width=0.4\textwidth]{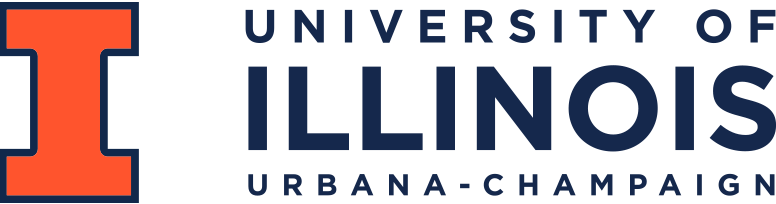}

        \vspace{0.8cm}
        \Large
        Department of Computer Science\\
        University of Illinois\\
        Urbana, IL\\
        May 12, 2023
            
    \end{center}
\end{titlepage}
\title{Attitudes and perceived effectiveness among first-time online instructors during Covid-19}

\begin{abstract}
Researchers have long worked to enhance access to quality education. Compared with traditional face-to-face instruction, online teaching provides pedagogical content to a larger audience in a more flexible education environment. However, while early studies have shed light on online teaching, it is crucial to understand the perspective of instructors who experienced the Covid-19 pandemic and subsequently conducted their first online classes, since the evolving landscape of post-pandemic teaching may bring new emphases and approaches. In addition, previous studies mainly focused on faculties who volunteered to teach an online course and may have been more open to the idea in the first place. In this study, my colleague Tiffany Wenting Li surveyed university faculties teaching online for the first time, regardless of whether they volunteered. She first surveyed them when the university began transitioning from in-person to online instruction in April 2020 due to the pandemic. Later, she conducted a follow-up survey as they completed their first online teaching semester. 

In the surveys, with my collegue Tiffany, we investigated instructors’ expected level of class success towards online teaching before their first online teaching experience. Using Bayesian modeling, we analyzed this expectation varied based on instructors’ characteristics (self-efficacy of online teaching, tech-savviness, and technology acceptance) and course features (subject area, class size, and instructional design). We found instructors' self-efficacy in their first online class had a significant impact on their expectations of success. Furthermore, our findings indicate that smaller class sizes are associated with lower expectations of success. On the other hand, instructional design factors such as physical space and prior use of technology platforms did not show significant contributions to the final expectation of success. Lastly, this study proposes practical recommendations to support online teaching for instructors and school administrators. To increase instructors' self-efficacy, they should collaborate with colleagues and familiarize themselves with online platforms. Universities should organize workshops or training sessions for instructors to enhance their skills. In small-size interactive classes, instructors should utilize nonverbal communication, while universities should establish support teams and implement feedback mechanisms to ensure quality and effectiveness.


\end{abstract}

\ccsdesc[500]{Computer systems organization~Embedded systems}
\ccsdesc[300]{Computer systems organization~Redundancy}
\ccsdesc{Computer systems organization~Robotics}
\ccsdesc[100]{Networks~Network reliability}

\keywords{datasets, neural networks, gaze detection, text tagging}


\maketitle

\section{Introduction}

Online teaching has a long growing history since 1990 \cite{martin2020systematic}. The goal of online teaching is to enhance access to education and address inequality \cite{britto2013three}. In recent years, online teaching has become increasingly prevalent, particularly during and after the pandemic. Previous studies \cite{mitchell2009attitudes, volery2000critical} have shown that instructors' attitudes and acceptance significantly impact their approach and performance in online teaching. For instructors who were teaching online for the first time, I and my colleague Tiffany Wenting Li, are concerned about their attitudes toward online teaching, specifically, their \textbf{expected level of success} of their first online class. It is important to note that previous studies may have been affected by self-selection bias, where instructors with more positive attitudes towards online teaching were predominantly included. The transition to online instruction during the Covid-19 pandemic has provided an opportunity to address this gap, as instructors who are teaching online for the first time may not be familiar with or favor online teaching. In this project, with my colleague Tiffany, we aimed to investigate the factors that influence the expected success level of online courses, considering both course features and instructor features. Based on our findings, we will propose recommendations to instructors to ensure quality and effectiveness in online teaching, as well as recommendations to school administrations to provide necessary resources and support to instructors. Since this is a collaborated work, in this thesis, the pronoun "we" collectively refers to both Tiffany Wenting Li and myself.

To study the features, we proposed 2 research questions in this thesis:

\begin{enumerate}
    \item[\textbf{RQ1}] For instructors without prior online teaching experience, what are their \textbf{Expected Level of Online Class Success} as they transform their in-person course to an online course?
    \item[\textbf{RQ2}] How does the \textbf{Expected Level of Online Class Success} vary based on Subject Area (IV1), Class Size (IV2), Course Design (IV3), Instructors' Self-efficacy of Online Teaching(IV4), Tech-savviness(IV5), Technology Acceptance(IV6)?
\end{enumerate}

To answer our research questions, we explored the factors that influence instructors' expectations of success before their first online class. Specifically, we investigated the impact of fields of study, class size, instructional design, self-efficacy of online teaching, tech-savviness, and technology acceptance. By conducting quantitative studies, we found that instructors' self-efficacy plays a crucial role in shaping their expectations of success. Additionally, we investigated the potential influence of class size and instructional design, such as physical space and technology used previously, on success expectations. Through our findings, we provide valuable insights into strategies for school administrations, including enhancing instructors' self-efficacy and explaining why smaller class sizes can be advantageous in the context of online teaching.




\section{Related Work}

\subsection{Selection bias and broad studies in online teaching}
The majority of related work on instructors' attitudes was conducted on instructors who were already teaching online courses. Such samples could suffer from a selection bias where these instructors were mostly volunteers who already held a more positive attitude than an average instructor. There are a few studies that investigated instructors including those who have not taught an online course before. Ward\cite{ward2008science} surveyed instructors in four STEM subjects about their attitudes toward online courses, while Vogle\cite{vogle2014instructors} surveyed instructors in Visual Arts. Corry and O'Quinn\cite{o2002factors}, Zhen et al. (2008), and Gasaymeh\cite{gasaymeh2009study} investigated factors that impacted an instructor's attitudes toward online courses and the decision of whether to teach an online course or not. The factors included self-efficacy of online teaching technology, perceived usefulness of online teaching, administrative support from the department, time-related challenges, personal experience with online courses, etc. However, these studies looked at their attitudes towards online teaching in general, but not in the context of a specific course they are teaching. Asking them in the context of a specific course 1) allows us to explore how course features (class size, subject area, etc.) impact attitudes, and 2) grounds instructors' responses so that the responses are less imaginary but more realistic.

\subsection{Novice professors of online teaching in the post-pandemic era}
While the pandemic eruption forced the online teaching transition, it also emphasizes the need for professional development and support in effectively delivering online instruction\cite{hodges2020difference}. For university instructors, previous studies showed the need for training and support to enhance their online teaching skills, which affect their decisions to teach online or not\cite{zhen2008factors, ward2008science}. When the transition is forced or highly recommended, how to make instructors, especially those who have not taught online classes before, willing to teach online became important. To give suggestions for first-time or inexperienced online teaching instructors, many previous studies\cite{schmidt2013university,lewis2006implementing, ulrich2011faculty} in the early 21st century have covered suggestions from authorities and experienced instructors. However, for instructors who are forced to teach online for the first time, suggestions from these prior studies may not be applicable to them, and also not up-to-date in the 2020s with post-pandemic teaching as they might have different emphases and strategies.

\section{Method}

My colleague Tiffany conducted two surveys for instructors in two universities before and after they taught their first online classes. In April 2020, most universities in the US switched from in-person instruction to online instruction. She recruited instructors from two universities that made the switch who had no online teaching experience before the switch. The survey was timed such that the instructors had no or very little online teaching experience when they filled in the first survey, and had taught half a semester online by the second survey. In the first survey, \textbf{pre-survey}, She collected 124 completed responses. After the Spring 2020 semester has ended, which was half a semester after the pre-survey, she sent the second follow-up survey, \textbf{post-survey}, to those who completed the pre-survey and agreed to be recruited for a post-survey, and collected 34 completed responses. In this thesis, she only focused on the pre-survey data to answer our research questions.

\subsection{Demographics}
Participants are instructors from 2 universities(UIUC, Uchicago) across diverse fields of study (Engineering, Business, Language, Natural Science, etc.). There are 124 participants in total. Among them, 59 are Women(47.6\%) , 62 are Men(50\%) and 3 are not-disclosed(0.24\%). And 76 are from UIUC(61.3\%), 46 are from the University of Chicago(37.1\%) and 2 refuse to answer(1.6\%). For the ages of participants, most of them are in the range of 25-75(98.4\%, 122 participants out of 124). Specifically, there are 1 participant aged below 25, 15 participants aged 26-35, 26 participants aged 36-45, 33 participants aged 46-55, 33 participants aged 56-65, 15 participants aged 66-75, and 1 participant aged over 75. 

In addition, from these 124 participants, two-thirds of them define "Professor" as their job title (101, 66\%), while the rest are Associate Professors(28, 22.6\%), Assistant Professors(19, 15.3\%), Lecturer(12, 9.7\%), Instructor(3, 2.4\%), Graduate Student(1, 0.8\%), and Others(6, 4.8\%). For their years of teaching experience in higher education, participants are concentrated in the 10-30 year interval. In detail, 11 participants(8.9\%) have experiences in less than 5 years, 17 participants(13.7\%) have experiences in 5 to 9 years, 32 participants(25.8\%) have experiences in 10 to 19 years, 31 participants(25\%) have experiences in 20 to 29 years, and 33 participants(26.6\%) have experiences in over 30 years. 

She also asked participants to describe the level of online teaching training they had received prior to their first online teaching experience. The results showed that many participants(26\%) had no training, and the number of participants in each training level decreased as the training level increased. From the collected responses, 48 participants(38.7\%) describe them as "No training at all", 32 participants(25.8\%) describe them as "Episodic informal training (e.g., reading a blog post)", 23 participants(18.5\%) describe them as "3 or fewer hours of formal introductory training over time", 12 participants(9.7\%) describe them as "More than 3 hours of formal introductory training over time", 2 participants(1.6\%) describe them as "3 or fewer hours of formal advanced level training over time", and 7 participants(5.6\%) describe them as "More than 3 hours of formal advanced level training over time".

Based on the department they are affiliated with, we grouped them into 6 fields of study. The following table shows the distribution by fields of study:
\begin{table}[H]
\begin{tabular}{|l|l|l|l|l|l|l|}
\hline
           & STEM & Arts and Humanities & Social Sciences & Public and Social Services & Health and Medicine & Business \\ \hline
Number     & 57   & 39                  & 14              & 6                          & 5                   & 3        \\ \hline
Percentage & 45.6\% & 31.5\%                & 11.3\%            & 4.8\%                       & 4.0\%                & 2.4\%     \\ \hline
\end{tabular}
\end{table}



\subsection{Pre-survey Design}
Our pre-survey included questions on the following themes: 
\begin{enumerate}
    \item Eligibility

    For the first two questions, the survey checked the participants' eligibility for this study. In the first question of the survey, the survey informed participants about the purpose of the study, which is to collect first-time online instructors' attitudes towards online teaching, and explain that their participation will involve a 12-18 minute online survey. In the second question, the survey asked if the university's teaching modality switch in April 2020 was their first time teaching a course fully online. Only if they certify the first question and select "yes" in the second question were they directed to the later parts.
    
    \item In-person Course Description
    
    First, the survey asked about the details of the course they were teaching in the following dimensions: learning objective, affiliated department, class size, targeted students, and grade options.
    
    \item In-person Course Component
    
    Then, the survey asked about the components involved in the in-person version of the course. The survey first asked, in the in-person version of the course, how much time was spent on instructor-student interaction and peer interaction in a typical lecture session, a typical lab or studio session, and a typical fieldwork or field practicum session respectively. Then, the survey asked what assessment methods, such as take-home exams and class participation, counted towards the final grade. It also asked what parts of the in-person course were conducted online and what technology artifacts or platforms were used in the in-person version of the course. 

    
    \item Initial Attitudes
    Next, the survey asked participants to look back and answer questions in this part of the survey as if they were just beginning to plan for the online course. The questions elicited their perceived expected usefulness of their online course, perceived expected ease of use for their online course, intention to teach their course online in the future, and their emotional attitudes towards online teaching.

    \item Demographic

    In this section, the survey asked for their personal information, including gender, age, teaching years, job title, level of training in online teaching, affiliated department, and affiliated university.


    \item Interest in a Follow-up Study

    At the end of the survey, the survey asked if they were interested in continuing to participate in the research study for a post-survey and to provide their emails if they wanted.

\end{enumerate}

\subsection{Variables of Interest}
Based on our research questions and the survey questions, we first defined variables of interest and their measurements. Then, we cleaned the collected responses and transformed them into the desired variable formats. We have three kinds of variables, Dependent Variables(DV), Independent Variables(IV), and Controlled Variables(CV). Our research questions ask about the effects of IVs on DVs, when controlling the confounds presented by the CVs.

\subsubsection{Dependent Variables}
Among the several attitude measurements in the survey, in this thesis, we focus on the \textbf{Expected Level of Online Class Success}, which is measured by the section of "the initial expectation of success" in the survey. This section included 16 questions. Among these questions, we asked about their expectations of the following aspects in the online version of the class compared to the in-person version of the same class: students' performance, flexibility, interaction, fairness, and quality.

\begin{table}[]
\centering
\label{tab:evaluation}
\begin{tabular}{|l|l|}
\hline
\multirow{4}{*}{Students' Performance} & Students' Learning Outcome \\ 
                                      & Students' Engagement and Participation \\ 
                                      & Students' Motivation \\ 
                                      & Amount of Help-seeking Behavior from Students \\ 
\hline
\multirow{3}{*}{Flexibility} & Flexibility of the Teaching Process \\ 
                             & Flexibility in Assessment and Evaluation \\ 
                             & Flexibility of the Learning Process for Students \\ 
\hline
\multirow{6}{*}{Interaction} & Interaction Time between Instructors and Students \\ 
                             & Interaction Quality between Instructors and Students \\ 
                             & Variety of Interaction Channels between Instructors and Students \\ 
                             & Interaction Time among Students \\ 
                             & Interaction Quality among Students \\ 
                             & Variety of Interaction Channels among Students \\ 
\hline
\multirow{1}{*}{Fairness}    & Fairness in Assessment and Evaluation \\ 
\hline
\multirow{2}{*}{Quality}     & Quality of Content Delivery \\ 
                             & Amount of Feedback Given to Students in Assessment and Evaluation \\ 
\hline
\end{tabular}
\caption{Evaluation Criteria}
\end{table}


The survey asked the above questions using a 7-point Likert scale, where 1 means the most successful and 7 means the least successful, compared with the in-person course. We calculated the average value in the question section of "the initial expectation of success" as the value of the \textbf{Expected Level of Class Success}. After transformation, we had 4.99 points as the average value, which shows participants had a slightly lower expectation of course success when it's taught online compared to in-person.

\subsubsection{Independent Variables}
For independent variables, we defined variables of interest in the following categories:
\begin{enumerate}
    \item Subject area(I1)
    
    The affiliated departments of the course. We deductively coded the answers into 6 subject areas: Arts and Humanities, Business, Health and Medicine, Public and Social Services, STEM, and Social Sciences.
    \item Absolute class size(I2a)

    The number of students in their classes.
    \item Relative class size(I2b)

    This refers to how big their class was relative to all the courses offered in their department. They have three options: small class, average-size class, and large class.
    \item Instructor-student interaction time(I3a)

    The survey asked how much time they used for interaction time between themselves and students throughout the class when the course was in-person. We calculated the average value of the corresponding answers and standardized them from 0 to 1, and then multiply this answer with 7 so that the outcome form is aligned with other 7-point Likert scaled variables. After these processes, we have the final value that ranges from 1 to 7, where 1 means the least time and 7 means the most time.
    \item Peer interaction time(I3b)
    
    Same with I3a, the survey asked how much time they used for interaction time among students throughout the class when the course was in-person. Since all the questions in this section are on a 7-point Likert scale, we calculated the average value of the corresponding answers that range from 1 to 7, where 1 means the least time and 7 means the most time.
    \item Physical space(I3c)

    The survey question asked about the necessary physical space (e.g., chemistry lab, dance studio) or equipment (e.g., lab equipment, mirrors) the course relied on when it was in-person. And we deductively coded their answers to physical space demands into the following categories: need special spaces, and no need.
    \item Technology(I3d)

    The survey asked them what technology artifacts (e.g., projector, iClicker) or technology platforms (e.g., Compass, Slack) were used when the course was in-person. And we deductively coded their answers of used technology into the following categories: both software and hardware, hardware only, software only, and no tech.
    \item Online components in the in-person course(I3e)

    The survey asked them what parts of the in-person course were conducted online. And we inductively coded their answers into the following categories: Only regular assignments/emails/forums were online versus more components were online.
    \item Self-efficacy towards teaching online(I4)

    The survey asked them several questions about their self-efficacy in online teaching. Since all the questions in this section are on a 7-point Likert scale, we calculated the average value of the corresponding answers that range from 1 to 7, where 1 means the most efficacious and 7 means the least efficacious.
    \begin{figure}[H]
        \centering
        \includegraphics[width=\textwidth]{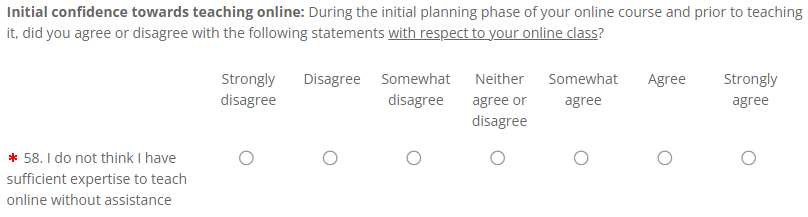}
        \caption{Example question for I4}
    \end{figure}
    \item Initial proficiency towards technology(I5)

    Similar to I4, the survey asked them several questions about their initial proficiency towards technology. And then we calculated the average value that ranges from 1 to 7, where 1 means the most proficient and 7 means the least proficient.
    \item Open attitudes towards technology(I6)

    Similar to I5, the survey asked them several questions about their open attitudes toward technology. And then we calculated the average value that ranges from 1 to 7, where 1 means the most openness and 7 means the least openness.
    \begin{figure}[H]
        \centering
        \includegraphics[width=\textwidth]{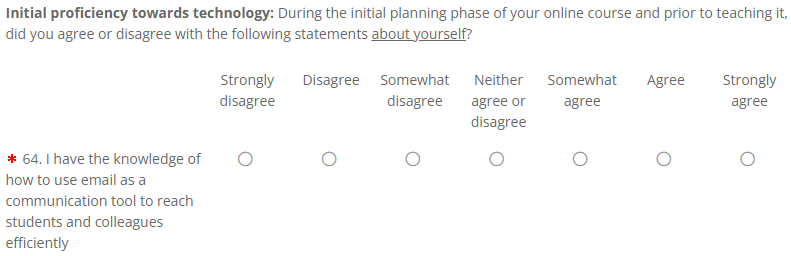}
        \includegraphics[width=\textwidth]{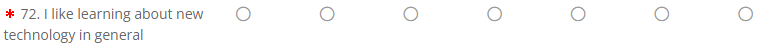}
        \caption{Example question for I5(\#64) and I6(\#72)}
    \end{figure}
\end{enumerate}

\subsubsection{Controlled Variables}


By applying the transformation rules, We transform the data into all Controlled Variables(CV) shown in the Table 2.


\begin{table}[]
    \centering
    
        \begin{tabular}{|l|l|}
        \hline
        \textbf{Variable} & \textbf{Name} \\ \hline
        C1 & Personal Experience with taking online courses (1: good, 7: bad) \\ \hline
        C2 & Prior impression about online teaching (1: good, 7: bad) \\ \hline
        C3a & The most important part in a course (content/peer-interaction/instructor-student interaction)\\ \hline
        C3b & In-person communication is always better (1: strongly agree, 7: strongly disagree) \\ \hline
        C3c & Unwillingness to change the style (1: strongly agree, 7: strongly disagree) \\ \hline
        C3d & Instructors self-defined role: facilitator (1: strongly agree, 7: strongly disagree) \\ \hline
        C3e & Instructors self-defined role: feedback-giver (1: strongly agree, 7: strongly disagree) \\ \hline
        C3f & Instructors self-defined role: expert (1: strongly agree, 7: strongly disagree) \\ \hline
        C4a & Gender \\ \hline
        C4b & Age \\ \hline
        C4c & Job title \\ \hline
        C4d & University \\ \hline
        C4f & Training in online teaching (0: no training, 5: most training) \\ \hline
        C5a & sufficient training (1: strongly agree, 7: strongly disagree) \\ \hline
        C5b & financial support (1: strongly agree, 7: strongly disagree) \\ \hline
        C5c & technology provided (1: strongly agree, 7: strongly disagree) \\ \hline
        C5d & tech issue assistance (1: strongly agree, 7: strongly disagree) \\ \hline
        \end{tabular}
    \label{tab:variables}
    \caption{Variable table}
\end{table}

\subsection{Directed Acyclic Graphs}
We build DAGs(Directed Acyclic Graph) to define causal relationships among these variables. Based on the features of DAGs, there are 3 possible relationships between variable A and variable B: no relations, A $\rightarrow$ B and B $\rightarrow$ A. We denoted the arrow as the direction of the causal relationship. For example, we believe there is a causal relationship in \textbf{I1: Subject area} $\rightarrow$ \textbf{I4: Self-efficacy towards teaching online}. If we change the subject area, the most important teaching part will also change, but not the reverse. We built the DAG for all variables and linked the variables that could have a causal relationship to other variables with directed arrows. The graph is shown in Fig.3.

\begin{figure}[]
    \centering
    \includegraphics[width=\textwidth]{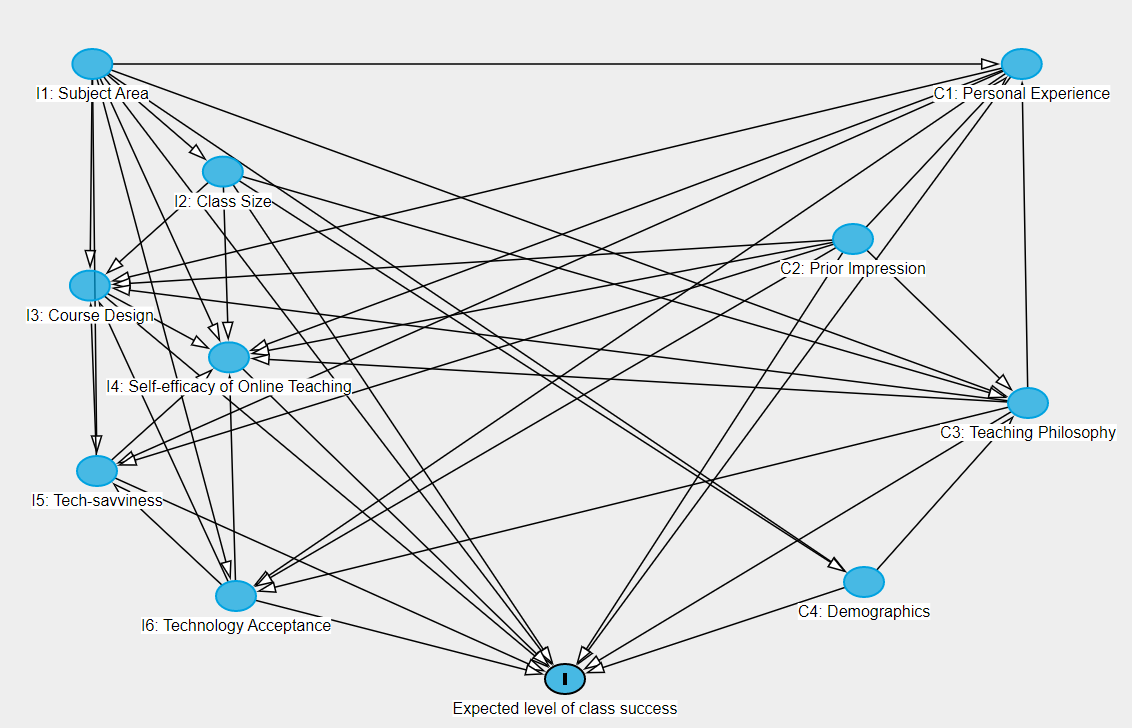}
    \caption{Directed Acyclic Graph of Expected Level of Class Success}
\end{figure}

\subsection{Bayesian Models}
We used Bayesian analysis to study the direct effects of each IV on our target DV, \textbf{Expected Level of Class Success}, under the circumstance that all other confounding variables (based on the DAG) in the model are controlled. We choose Bayesian analysis as the main tool for our quantitative study due to the following reasons\cite{xiao2023inform, cheng2021can}: 

First, since we only have less than 200 data points, Bayesian methods can help to compensate for the limitations of small sample size data\cite{gelman2013bayesian, mcelreath2020statistical} and improve the analysis accuracy with prior information or beliefs about the data. 
Second, Bayesian methods tend to be more flexible and robust than traditional frequentist methods in the context of small sample sizes, as they do not rely on strict assumptions about the distribution of the data. 
Third, Bayesian models offer the advantage of allowing for explicit specification and incorporation of all aspects of the model, without requiring the need to check modeling assumptions that are not already included in the model description, thereby foregrounding all relevant assumptions and parameters within the model.
Lastly, In contrast to null hypothesis significance testing (NHST), Bayesian analysis prioritizes quantifying the strength of an effect rather than simply determining its presence, which aligns more closely with the exploratory nature of many studies. As a result, Bayesian analysis can provide more informative and nuanced insights into the data under investigation.

To better explain what a direct effect of an IV on a DV is, let's look at an example of the following causal relationships: I1 $\rightarrow$ I3 $\rightarrow$ D1; I1 $\rightarrow$ D1, we included I3 in the model of studying I1 $\rightarrow$ D1 so that we can block the effect of I3 $\rightarrow$ D1, and hence we can study the direct effect of I1 $\rightarrow$ D1. Based on the DAGs we built, to estimate the direct effect of each IV to the DV, we carefully choose which IVs and CVs to include in each model such that all the confounds were taken care of and no new confounds were created. We created a total of 8 different models. 

After plotting the histogram of our target outcome variable \textbf{Expected Level of Class Success}, we found it would fit the Normal distribution. So we modeled the data as a Normal distribution and used linear regression models to estimate the Normal distribution means and standard deviation for different values in the exposure variables. 

To calculate how the DV changes per unit change in an IV, we took different approaches for a categorical IV and a Likert-scale continuous IV. We define a unit change of a category IV as changing from one category to another, and a unit change in a Likert-scale IV as changing one Likert level. So for categorical ones, we contrast each two of all possible categories; for continuous ones, we contrast them with a unit point difference in a 7-point Likert scale. By contrasting the posterior distributions of the means for 2 different categorical conditions or 1 point difference in the 7-point Likert Scale, we can see how the exposure variable affects outcome variables. And we also estimated the posterior distribution of Cohen's D for the linear regression model as a measurement of relative effect size.

\section{Result}
Regarding our RQ1(For instructors without prior online teaching experience, what are their \textbf{Expected Level of Online Class Success} as they transform their in-person course to an online course?), the data showed that participants have slightly lower expectations of success(Mean: 0.665, Standard Deviation: 0.129), compared to the in-person version of the same class. Mean value 0.665 approximates the 5th point on a 7-point Likert scale, corresponding to slightly lower success. In the following subsections, we will dive into how course features and instructors' features impact the level of expected success, in answer to our RQ2.

\subsection{No significant direct effects from subject to the expected level of success}
For all the responses we received, we grouped all the responses in "subject area" into 6 categories, which are: Arts and Humanities, Business, Health and Medicine, Public and Social Services, STEM, and Social Sciences. And we are curious how each subject area differs from others in the \textbf{Expected Level of Class Success}. Among these six categories, Art and Humanities(size:39, Mean:0.655, Standard Deviation: 0.125), STEM(size:57, Mean:0.652, Standard Deviation: 0.140), and Business(size:3, Mean:0.633, Standard Deviation: 0.073) have higher expectation levels(Mean: 0.646); while Public and Social Services(size:6, Mean:0.725, Standard Deviation: 0.126), Health and Medicine(size:5, Mean:0.71, Standard Deviation: 0.099), Social Sciences(size:14, Mean:0.708, Standard Deviation: 0.709) have lower expectation levels(Mean: 0.714): the mean value of the latter three categories is 0.5 points less than the former three categories in 7-point Likert scale. After controlling for other confounds(I2, I3, I4, I5, I6, C1, C2, C3, C4) to be the same values, we modeled the direct effect of \textbf{I1: Field of study} as a categorical variable on the \textbf{Expected Level of Class Success}. We contrasted the posterior distributions of \textbf{Expected Level of Class Success} of each subject area against the average of other subject areas. For each subject area, the HPDI(High-Posterior Density Interval) overlapped with the value of 0. This indicated that given the same value of confounds\textbf{(}same personal experience ratings, prior impression about online teaching ratings, teaching philosophy, gender, age, job title, university, training, class size, interaction time, physical space needs, technology used, self-efficacy towards teaching online, initial proficiency towards technology, and openness to technology, self-efficacy towards teaching online\textbf{)}, simply being in different fields of study did not have any significant direct effect on the initial expectation of success. The mean absolute differences of each category to the average of other categories are small (absolute value less than 0.01). 


\subsection{Self-efficacy affects success expectations}
We find that instructors' \textbf{Self-efficacy towards Online Teaching(I4)} has a significantly positive direct effect on \textbf{Expected Level of Class Success}.

Among all participants, we received overall good self-efficacy ratings(Mean: 0.384, Standard Deviation: 0.180), where 0 means the most efficacious and 1 means the least efficacious. 0.384 means approximately 3.3 points on a 7-point Likert scale, which is somewhat agreeing they are efficacious.

As a continuous variable,  we consider how much each unit of \textbf{Self-efficacy towards Online Teaching(I4)} change would affect the target variable. We contrast the posterior distribution of \textbf{Expected Level of Class Success} when I4 is 0.5 and 0.667 (subtract 0.5 group from 0.667 group), where the difference is 1 unit point in a 7-point Likert scale. From the plotted graphs, we found that the HDPI range(Absolute Difference: [0.017, 0,0655], Effect Size: [0.25, 1]) excluded the reference value of 0. The effect size is on the right side of value 0, which shows the effect is positive. To conclude, there is a significant effect from \textbf{Self-efficacy towards Online Teaching(I4)} to \textbf{Expected Level of Class Success}. This means that if participants have more self-efficacy toward online teaching, they would have higher expectations for success in class. The graph is shown in Fig.4.

\begin{figure}
    \centering
    \includegraphics[width=\textwidth]{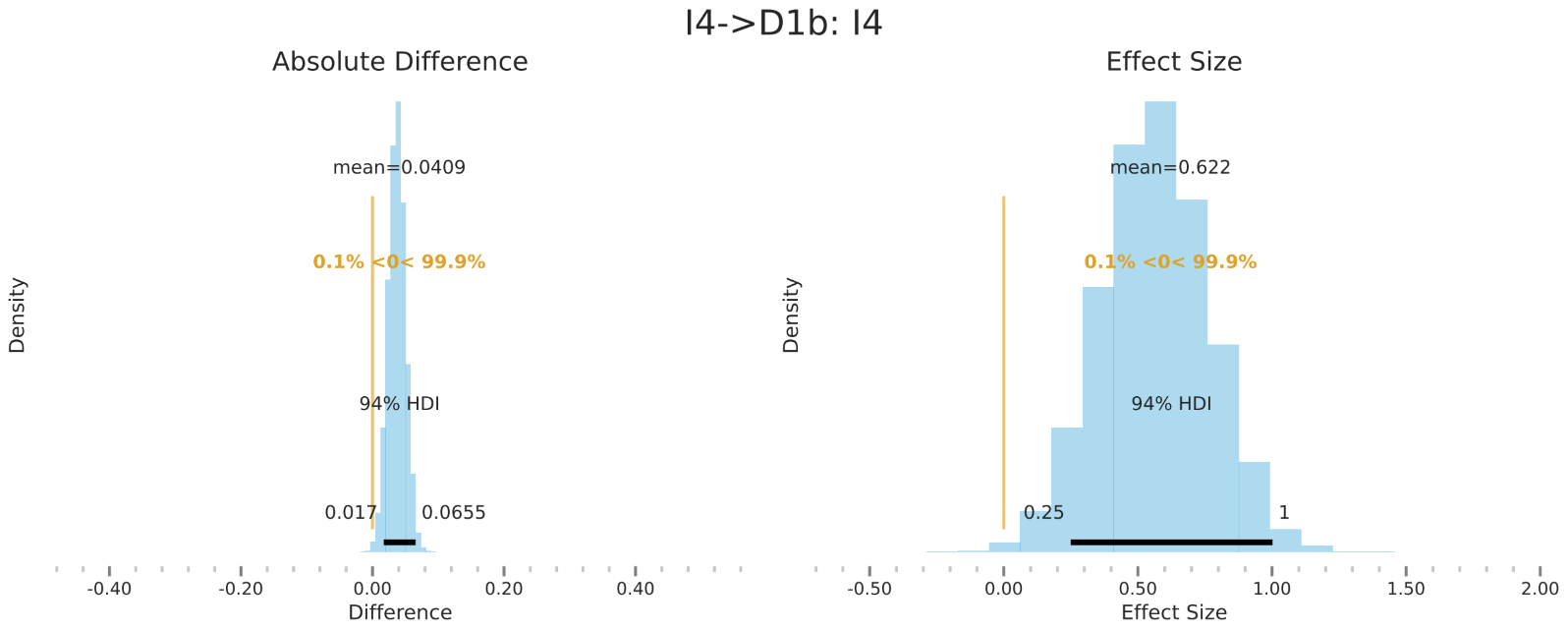}
    \caption{Posterio distribution of \textbf{Expected level of Class Success} varied by changing I4(subtract 0.5 group from 0.667 group)}
\end{figure}

\subsection{Best class size for online teaching?}
To study what would be the perfect size for online instruction, in other words, in what size the instructors feel the most successful, we asked their \textbf{Absolute Class Size(I2a)} and \textbf{Relative Class Size(I2b)}, where we defined them in Section 3.3.2. Regarding the class size, small-sized classes (number of responses: 33, Mean: 0.703, Standard Deviation: 0.145) are considered to have a lower(0.35 and 0.26 points lower on the 7-point Likert scale) success level than average-sized ones (number of responses: 57, Mean: 0.645, Standard Deviation: 0.115) and large-sized (number of responses: 34, Mean: 0.660, Standard Deviation: 0.131). 

For the continuous variable, \textbf{Absolute Class Size(I2a)}, though the result HPDI of absolute difference ([-0.043, 0.0116]) and effect size ([-0.654, 0.172]) overlapped with the reference value of 0, a large proportion of the intervals concentrates on the left-hand side of zero (Left: 86.7\%, Right: 13.3\%). This shows a trend that as the number of students in a course increases, the expected level of class success would be higher. The graph is shown in Fig.5.

\begin{figure}
    \centering
    \includegraphics[width=\textwidth]{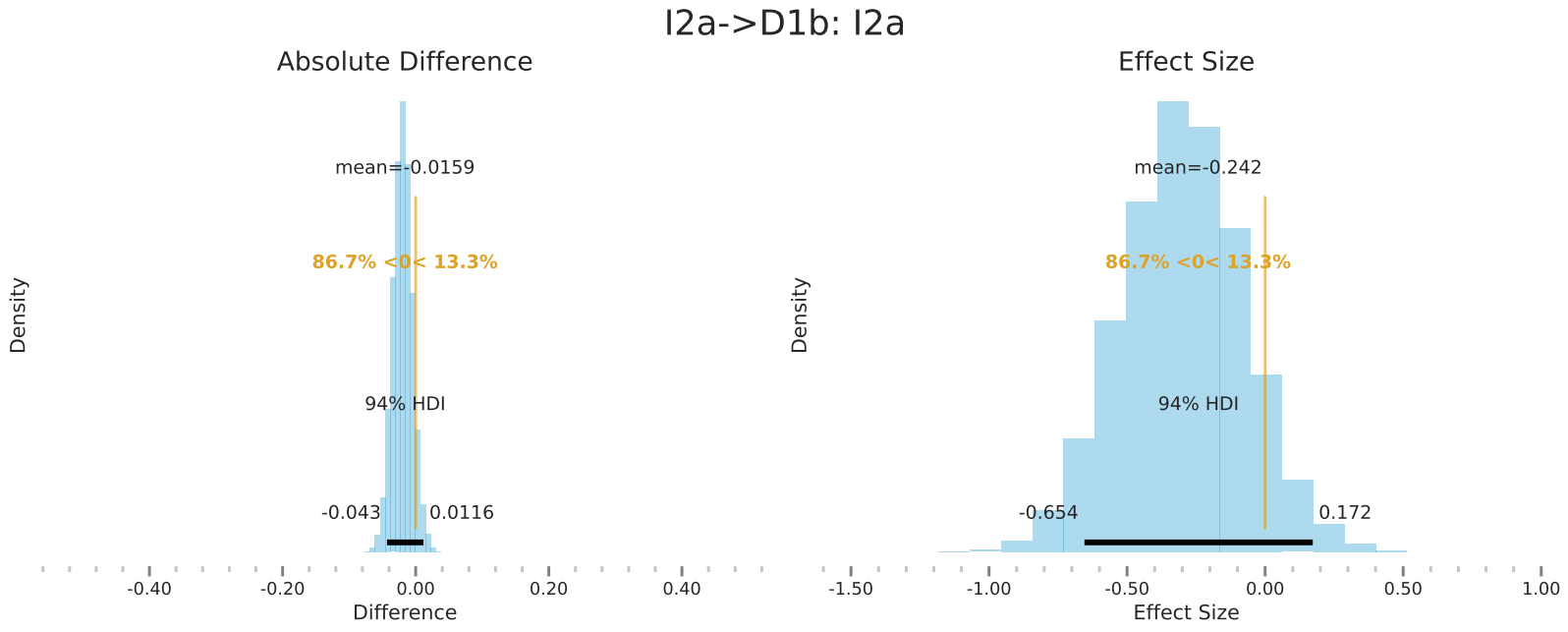}
    \caption{Posterio distribution of \textbf{Expected level of Class Success} varied by changing I2a(subtract 0.5 group from 0.667 group)}
\end{figure}


Results for the relative class size (I2b) showed similar trends. When comparing large class sizes to small class sizes, the absolute difference in expected class success rating ranges from -0.083 to 0.0239, and the effect size ranges from -0.809 to 0.206. This suggests that, on average, teaching large-sized classes online tends to have slightly higher levels of expected success compared to small-sized classes. Similarly, when comparing average class sizes to small class sizes, the absolute difference in expected class success rating ranges from -0.087 to 0.011, and the effect size ranges from -0.839 to 0.098. This implies that, on average, average-sized classes also tend to have slightly higher levels of expected success than small classes.

However, when comparing average class sizes with large class sizes, the data shows that the mean difference is -0.0588, which is very close to zero. This means that there is no significant difference in the expected levels of class success between large-sized and average-sized classes.

In summary, instructors of smaller classes showed lower expectations of success, while there is no substantial difference in the expected levels of class success between large-sized and average-sized classes. We will dive into this in the discussion section.


\subsection{Insignificant direct effects from other IVs}
We did not find any significant or near-significant direct effects for other independent variables to \textbf{Expected Level of Class Success}. 

We concluded that instructor-student Interaction(I3a) (HDPI: Absolute Difference: [-0.019, 0.013], Effect Size: [-0.277, 0.21]), Peer interaction(I3b) (HDPI: Absolute Difference: [-0.02, 0.01], Effect Size: [-0.308, 0.146]), Physical space(I3c) (HDPI: Absolute Difference: [-0.074, 0.041], Effect Size: [-0.662, 0.372]), Technology(I3d), online components(I3e), Initial proficiency towards technology(I5) (HDPI: Absolute Difference: [-0.026, 0.012], Effect Size: [-0.391, 0.174]), and Open attitudes towards technology(I6) (HDPI: Absolute Difference: [-0.021, 0.018], Effect Size: [-0.317, 0.269])have no significant effects by checking their HPDI range. We will also discuss how these results compare to our hypotheses in the discussion section.

\section{Discussion}

\subsection{Building confidence to expect success}
In Section 4.2, the findings reveal a significant positive causal relationship between higher \textbf{Self-efficacy toward Online Teaching (I4)} to increased expectations for success. This outcome supports our initial hypothesis. The direct effects from \textbf{initial proficiency towards technology (I5)} and \textbf{open attitudes towards technology (I6)} to the outcome measurement did not demonstrate statistical significance, which contradicted our hypotheses and previous studies results\cite{phan2017teacher, gasaymeh2009study}. Our best explanation for this contradiction is that I5 and I6 affect success expectations through I4. In support of this explanation, we observed notable correlations between I5 and I4, as well as between I6 and I4, with Pearson correlation coefficients of 0.53 and 0.47\cite{cohen2009pearson}, respectively. These results suggest a moderate positive association between I4 and both I5 and I6. Therefore, we posit that enhancing instructors' open attitudes towards technology, improving their proficiency in technology, and fostering their self-efficacy are crucial considerations for promoting success. Below, we propose recommendations that focus on strategies to cultivate instructors' open attitudes, enhance their technological proficiency, and bolster their confidence levels.

First, we encourage instructors to collaborate with colleagues with experience teaching online. Experienced colleagues can provide guidance, share best practices, and offer support during their transition. Collaborating with experienced online instructors provides valuable insights and helps build confidence through shared knowledge and experiences. Second, instructors can enhance their online teaching experience by familiarizing themselves with the online platform they will be using. It is highly recommended for instructors allocate time to explore the platform's features and functionality. By doing so, they can become comfortable with the platform's interface, tools, and options. This familiarity allows them to navigate through the platform effortlessly and access the necessary teaching resources with ease.

On the other hand, universities should offer online teaching workshops or training sessions specifically designed to help instructors transition to online teaching, especially for those who have lower self-efficacy in online teaching. These sessions can provide valuable guidance on effective online teaching strategies, technology usage, and engagement techniques. Participating in these workshops can boost instructors' confidence and provide them with the necessary skills to teach effectively online.

\subsection{Small size class in online teaching}
Prior to the data collection, we believe online teaching brings more flexibility and compatibility in containing larger classes while inhibiting in-person interaction in small group classes, which would make large classes more successful than smaller ones. In section 4.3, we found our results matched the hypothesis: smaller size classes have lower success expectations, compared with average size and larger size. At the same time, for small-size classes, both peer interaction time and peer interaction time are higher than average classes and large classes. Even after the transition, small classes still kept the trends of emphasizing in-class interaction. In other words, although small-size classes contained relatively more intensive interaction, online teaching limited their expected success. Hence, we propose recommendations for universities and instructors to practice in small-size interactive classes.

First, instructors can use more visual cues and nonverbal communication during class. Online platforms with video capabilities allow for visual cues and nonverbal communication, helping instructors gauge student understanding, reactions, and engagement. Similarly, students can observe and learn from their peers' nonverbal cues, enriching their learning experience even in an online setting. Second, instructors can manage time for more one-on-one support. Online platforms offer tools like private messaging and quick responses, allowing instructors to give individual attention to students. In smaller online classes, instructors can closely track student involvement, offer personalized assistance, and make sure students grasp the material and make progress.

For departments and universities, first, it is beneficial to establish online learning support teams that include instructional designers, technology specialists, and online learning experts who can assist instructors in designing interactive online activities, troubleshooting technical issues, and providing guidance on best practices. Second, universities can enhance the quality of online teaching by conducting regular evaluations and gathering feedback at both the campus-wide and department-wide levels. By implementing mechanisms for collecting feedback from instructors and students, universities can gain valuable insights into their online teaching and learning experiences. This feedback can then be utilized to identify areas for improvement, address challenges, and refine online teaching strategies. Additionally, conducting evaluations of online courses can ensure ongoing quality assurance and effectiveness in the delivery of online education.

\subsection{Instructional design}

In Section 4.4 of our study, we examined the impact of various instructional design variables on the expected success level of online teaching. Specifically, we focused on variables related to physical space demand, technology utilization, and the inclusion of online components (I3c, I3d, and I3e, respectively). According to our initial hypothesis, classes that required specific physical spaces, both hardware and software use, and more online components would encounter difficulties in their initial online teaching experience.

However, our analysis revealed that these variables did not have a significant direct effect on the \textbf{Expected level of class success}. While they did show a tendency towards lower success expectations, their effects did not reach statistical significance. But this does not mean these variables do not have a significant total effect on expected success. It is worth noting that we believe these variables causally affect another variable, I4, which significantly affects \textbf{Expected level of class success}. As I4 is already incorporated into the model, it is possible that its inclusion "weakened" the effects of the aforementioned variables, thereby diminishing their direct impact on success expectations.


\section{Future Work}
\begin{enumerate}
    \item Other attitude measurements 
    
    In this thesis, we only considered one dependent variable: \textbf{Expected level of class success}. Though we have found some interesting results from \textbf{Expected level of class success}, we wonder what are other dependent variables like, what independent variables would have a significant effect on them, and what differences or similarities they would show with \textbf{Expected level of class success}. In the future, we will explore more interesting dependent variables like instructors' perceived expected usefulness of their online course, perceived expected ease of use for their online course, intention to teach their course online in the future, and their emotional attitudes towards online teaching.
    \item Changes in attitudes after teaching online
    
    For the data source, we did not include the post-data, so we do not know what changes have been made after these participants' first-time online teaching experience. In future work, we would also include post-data and investigate how DVs towards the online version of a course changed by the end of the course, and how DVs in post-data vary based on that DV in pre-data and IVs.
    \item Nuances in open-ended free-form answers
    
    Lastly, we only considered selected questions(mostly multiple-choice questions) in the pre-survey, which means that we missed some important open-ended free-form answers. We believe these free-form opinions are interesting and may explain more about the reasons and thought processes behind the answers. We will use those answers in future work for further qualitative studies to seek a broader understanding for instructors who teach online for the first time. 
\end{enumerate}



\section{Conclusion}
In this thesis, we conducted quantitative studies to investigate the expectations of first-time online teaching instructors before their transition. Our research examined various factors such as fields of study, class size, instructional design, self-efficacy of online teaching, tech-savviness, and technology acceptance. Through our findings, we discovered that instructors' expectations of success in their first online class were significantly influenced by their self-efficacy toward online teaching. Additionally, we found that smaller class sizes were associated with lower expectations of success, while factors such as instructional design elements and prior technology platform usage did not significantly impact the final expectation. Based on our research, we have provided recommendations for instructors and universities to build self-efficacy and ensure quality and effectiveness in online teaching. Additionally, we have provided suggestions specifically tailored for small-size interactive classes in the online teaching context.

\section*{Acknowledgement}

I would like to express my sincere gratitude to my collaborator Tiffany Wenting Li for her invaluable assistance in the data sharing, planning, tutoring, and revising of my thesis. Additionally, I would like to extend my thanks to my thesis advisor Prof. Karrie Karahalios for her invaluable guidance throughout the process, as well as Prof. Hari Sundaram for his assistance. I am also deeply grateful to all the survey participants for generously sharing their insights. Last but not least, I would like to acknowledge the unwavering support and love of my family and friends, who have been my pillars of strength throughout this journey.

\bibliographystyle{ACM-Reference-Format}
\bibliography{ref}










\end{document}